# A Formal Method For Mapping Software Engineering Practices To Essence

Murat Pasa Uysal

Department of Management Information Systems, Başkent University, Ankara, Turkey

## ABSTRACT

*Essence Framework (EF) aims at addressing the core problems of software engineering (SE) and its practices. As a relatively new framework, one important issue for EF has been mapping software practices to its conceptual domain. Although there are several works describing systematic procedures, a review of literature cannot suggest a study using a formal method. The study is conducted according to the guidelines of Design Science Research (DSR) Method. The research contribution is classified as an "application of a new solution (the formal method) to a new problem (mapping software practices to EF). The formal method employs an algorithm based on Concept Algebra and it is applied in a Scrum case study. The results are promising and they differ from the ones exist in the current EF related literature.*

## 1. INTRODUCTION

Software Engineering (SE) may be regarded as a relatively young discipline when compared to other disciplines from an evolutionary point of view. Technical innovations changing from time to time have been a major driving force for SE trends and practices. It has been usually driven by industrial needs, and thus, language-centred computer programming has been dominant in SE. However, the fundamental problems in SE exist today. SE industry still faces the major problems despite the developments in the methods, models, tools and techniques of SE knowledge domain. The Essence Framework (EF) is proposed for addressing the core problems of software development (SD) and its application [1]. Existence of plenty of development methods, which are: (a) hard to compare, (b) lacking of sound experimental method evaluations and/or validations; and (c) the increase of gap between practical application and academic research would be some of these problems. EF Kernel and Language Specification describes its key features and how it supports practitioners and method engineers. A set of elements for forming a common ground and describing a software engineering (SE) endeavour is defined as the kernel. Therefore, EF allows "people to describe the essentials of their existing and future methods and practices so that they can be compared, evaluated, tailored and re-used by practitioners as well as academics and researchers [2]".

By applying the principle of separation of concerns, and separating the "what" of SD from the "how", EF provides a common base and enables method building with the composition of various practices. Thus, a practice is defined as "a repeatable approach to doing something with a specific objective in mind [2]". It includes the necessary elements that exist in every software endeavour, such as, team work, requirements analysis/specification, development, test etc. Therefore, a method is built by the composition of a set of practices and using Kernel specifications.

EF includes a layered architecture with three discrete areas of concern. Each focuses on core and specific aspects of SE practices: (a) Customer, (b) Solution, and (c) Endeavour areas as depicted by Figure 1. In fact, the much of focus is given on the SD and practice use for compositing SD methods. The Alpha(s) of EF and the agile approach adopted enable capturing the key SE concepts. On this common ground, they allow monitoring the health and progress of SE endeavours and their associated artefacts. One of the key features of EF is that it allows a project team to assemble the methods according to their needs and experiences by the composition of various practices. However, an important issue has been how to map a SE practice to EF knowledge domain.

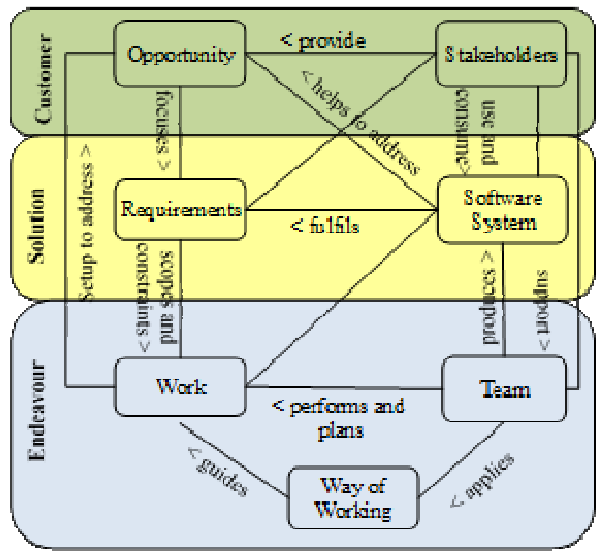

Figure 1. Essence Framework [1].

There are several works describing or proposing systematic translation of SE practices to EF-based descriptions. Essence Specification Document [2] includes several practice definitions, however, it has only a limited number and it is mainly for descriptive purposes. Park et al. base their mapping procedures on activity spaces, and thus, they propose an activity-state mapping algorithm, and present it in an Essence-powered Scrum practice [3]. Both Park [4] and Giray et al. [5] proposes an ontology-based systematic method for mapping SD to the EF. It is also explained how method engineering can help resolve some of the mapping issues [5]. In another study, Genetic Algorithms are introduced to generate candidate Essence Kernel replacements based on empirical data rather than human experience and judgement [6]. However, a review of literature on EF cannot suggest a formal method that guides mapping a SE practice to Essence-based definitions [7].

In this study, we propose a formal method for mapping SE practices to EF based on an algorithm using Concept Algebra definitions [8]. The next parts include theoretical foundations, method, sample case and conclusion sections respectively.

International Journal of Software Engineering & Applications (IJSEA), Vol.9, No.6, November 2018

## 2. THEORETICAL FOUNDATION

### 2.1. SOFTWARE DEVELOPMENT PROCESS

The Institute of Electrical and Electronics Engineers defines SE as "the application of a systematic, disciplined, quantifiable approach to the development, operation, and maintenance of software". It is an applied discipline and it encompasses processes, methods, tools, standards, and principles in order to build reliable, maintainable and large-scale software systems with high productivity and quality.

In the same context, software development process is the combination of a set of interrelated activities functionally coherent and reusable for SE. It transforms various inputs into outputs by using resources, tools and techniques. SE processes and activities (planning, requirement, design, construction, testing, and maintenance) occur both at the organizational and project levels. They ensure that the software product is delivered for the benefits of all stakeholders effectively and efficiently. Therefore, SE methods and the use of information, behavioural, structured modelling techniques provide a systematic approach to both problem solving and software development. Software processes are included in Software Development Life Cycle (SDLC), and it can be linear, iterative or incremental.

SDLC begins with the software requirement process that includes elicitation, analysis, specification, validation and management of the needs and constraints placed on a software product and its development. Software design process consists of architectural and detailed design, which in turn describes how software is organized into components and desired behaviours in sufficient detail. As one of the life cycle stages, the design process produces a description of the software's structure with a set of models and artefacts serving as the basis for its construction. Software design and architecture methods provide a common framework for software engineers.

### 2.2. CONCEPT ALGEBRA

Modelling is a kind of knowledge representation, and thus, conceptual mapping and semantic evaluations usually require formal methods. Since mapping the SD concepts of any SE practices to the EF concepts cannot be straightforward, thus, core concepts from Essence are initially extracted, and then, the mapping is conducted based on the formal definitions of Concept Algebra (CA) [8]. This algebra is "an abstract mathematical structure for the formal treatment of concepts and their algebraic relations, operations, and associative rules for composing complex concepts [8]". It mainly provides denotational mathematics principles for algebraic manipulations of concepts.

A concept is defined as "a cognitive unit to identify and/or model a real-world concrete entity or a perceived-world abstract subject [8]". Accordingly, a concept connotes attributes or properties, and it denotes members or instances. Compositional and relational operations are the two main operations of CA. Thus, problems of various knowledge domains, such as, software and system engineering, can be identified, manipulated and modelled by using CA operations. In this study, the relational operations are used for comparing and mapping the corresponding abstract concepts of a SE practice to the semantic context of EF ($\Theta$).

Given that Θ is a semantic context, the main conceptual definitions are as follows:

$$\Theta = (O, A, R) \tag{1}$$

Where, the symbol $O$ denotes a finite/infinite nonempty set of objects, $A$ is a finite/infinite nonempty set of attributes, and $R$ is a set of relations between $O$ and $A$. The general structured model of an abstract concept is illustrated in Figure 2.

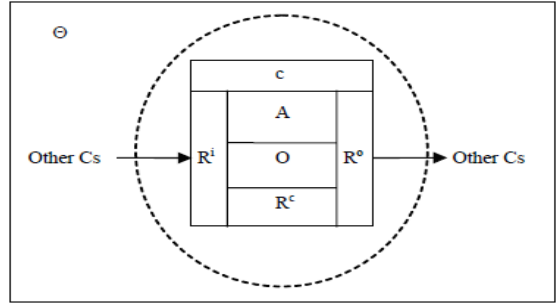

Figure 2. The structured model for an abstract concept [7]

An abstract Essence concept is regarded as the composition of different elements. Thus, an EF concept, with its attributes and objects, internal and external relations, can be defined as follows:

$$C_{EF} = (O_{EF}, A_{EF}, R_{EF}^{c}, R_{EF}^{i}, R_{EF}^{o}) \tag{2}$$

Where,
- $C_{EF}$ is a concept in Essence,
- $O_{EF}$ is a non-empty set of objects extended from this Essence concept, $O_{EF} = \{o_1, o_2, ..., o_m\}$,
- $A_{EF}$ is a non-empty set of attributes of EF objects, $A_{EF} = \{a_1, a_2, ..., a_n\}$,
- $R_{EF}^{c} = O_{EF} \times A_{EF}$ is a set of internal relations of the Essence concept,
- $R_{EF}^{i} \subseteq C' \times C_{EF}$ is a set of input relations of the Essence concept and where $C'$ is a set of external concepts,
- $R_{EF}^{o} \subseteq C_{EF} \times C'$ is a set of output relations.

A corresponding abstract SE Practice (SEP) concept, $C_{SEP}$, can be defined by adopting the same approach:

$$C_{SEP} = (O_{SEP}, A_{SEP}, R_{SEP}^{c}, R_{SEP}^{i}, R_{SEP}^{o}) \tag{3}$$

The relational operations in CA are defined as "related", "independent", "sub-concept", "super-concept", "equivalent", "consistent", "comparison", and "definition"; and they are represented by the $\{\mapsto, \leftrightarrow, \prec, \succ, =, \cong, \sim, \triangleq\}$ symbols respectively. Thus, the relationships between two concepts in the knowledge domains of EF and SEP are determined by the relations of their set of attributes $A$ and the set of objects $O$. As being a dynamic mathematical structure, it is important to note that an abstract concept can adapt and interrelate itself to other concepts via input relations $R^i$ and output relations $R^o$. In this study, these are $R_{SEP}^i$-$R_{SEP}^o$ and $R_{EF}^i$-$R_{EF}^o$ respectively.

## 2.3. DEFINITIONS

Take the concept $c_1$ from EF $\Theta$ and the concept $c_2$ from a SEP $\Theta$. Suppose that they have the sets of attributes $(A_1, A_2)$ and the sets of objects $(O_1, O_2)$. The following definitions are used when finding the similarity of two concepts in SEP and EF:

***Definition 1:*** See whether the related concepts $c_1$ and $c_2$ share some common attributes in $A_1$ and $A_2$, which are denoted by:

$$c_1 \leftrightarrow c_2 \Rightarrow A_1 \cap A_2 \neq \emptyset \tag{4}$$

***Definition 2:*** Compare $c_1$ and $c_2$ and determine their equivalency or similarity levels as below:

$$c_1 \sim c_2 \Rightarrow \frac{\#(A_1 \cap A_2)}{\#(A_1 \cup A_2)} * 100\% \tag{5}$$

Where # means the cardinal operator giving the number of elements in a given set, and thus, 0% means no similarity whereas 100% means a full similarity.

***Definition 3:*** Assume the equivalent concepts as follows:

$$c_1 = c_2 \Rightarrow (A_1 = A_2) \land (O_1 = O_2) \tag{6}$$

Which means that these two concepts have similar attributes $(A_1 = A_2)$ and their instances are identical $(O_1 = O_2)$.

## 3. METHOD

The study is conducted by following the guidelines of Design Science Research (DSR) [11]. This research method focuses on the creation of scientific knowledge when solving a real-world problem and developing IT artefacts in the Information Technology (IT) knowledge domain [7]. Constructs, models, methods and instantiation of a theory or solution are the main research outputs of a DSR project. (a) Building constructs, a model, a method or an instantiation; (b) evaluation of the quality of what was built; (c) theorizing about the quality and deciding whether it is satisfactory, and finally, d) justifying what was theorized are its main cyclic research activities (Fig 2).

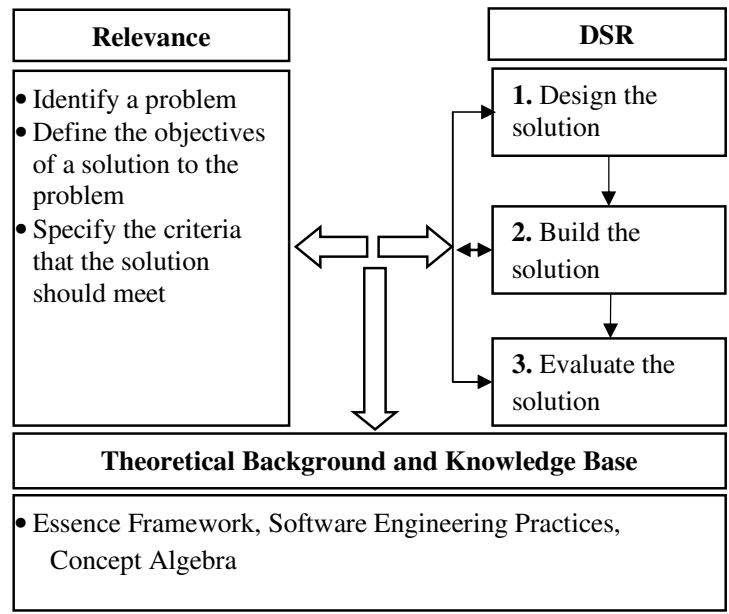

Figure 2. DSR Method

The research context and theoretical framework rely on the foundations of Software Engineering, Essence Framework and Concept Algebra. In terms of research contributions, the main research output can be classified as (a) developing a formal method as a new solution and then (b) applying this solution to a new problem (mapping software engineering practices to Essence) [12]. Tables or lists, which include both SEP and EF concepts, are initially created by using expert judgment method. And then, the proposed algorithm is applied: (a) to find whether the concepts share common attributes; (b) to calculate similarity levels and (c) to form a corresponding concept list of the SEP for EF (Fig 3).

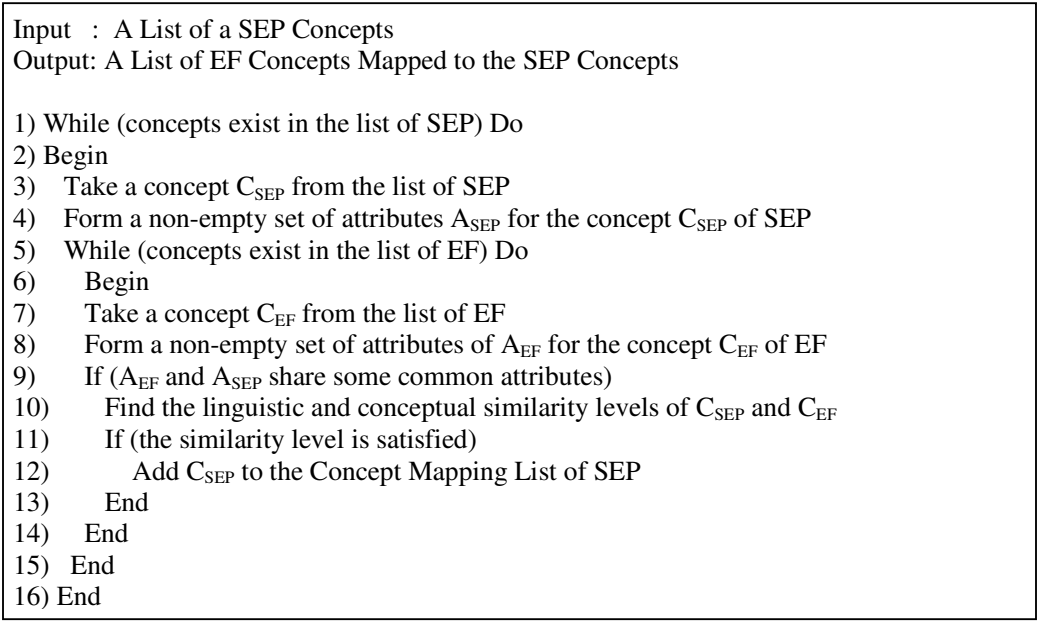

```
Input   : A List of a SEP Concepts
Output: A List of EF Concepts Mapped to the SEP Concepts

1) While (concepts exist in the list of SEP) Do
2) Begin
3)    Take a concept C_SEP from the list of SEP
4)    Form a non-empty set of attributes A_SEP for the concept C_SEP of SEP
5)    While (concepts exist in the list of EF) Do
6)       Begin
7)       Take a concept C_EF from the list of EF
8)       Form a non-empty set of attributes of A_EF for the concept C_EF of EF
9)       If (A_EF and A_SEP share some common attributes)
10)         Find the linguistic and conceptual similarity levels of C_SEP and C_EF
11)         If (the similarity level is satisfied)
12)            Add C_SEP to the Concept Mapping List of SEP
13)         End
14)      End
15)   End
16) End
```

Figure 3. Algorithm for mapping software engineering practices to Essence Framework

## 4. CASE STUDY

One of the well-known practices is the illustration of how Scrum [10, 11] can be modelled in the Essence Kernel and Language Specification [2]. In this document, "Product Backlog" concept of Scrum is associated with the "Requirements" alpha concept of EF without specifying conceptual details. Note that a comprehensive comparison of concepts exists in Scrum, and mapping them to EF is beyond the scope of this paper. However, it is thought that even in a simple and clear case, such as "Requirements" and "Product Backlog", it is possible to miss or neglect some important conceptual details. Therefore, the below section shows how the formal mapping is applied:

- The theoretical background of mapping is based on Concept Algebra principles and definitions.

- A content analysis for the EF specification document and resources related to Scrum Practice [2, 7, 8, 9] is conducted.

- An attribute comparison list is created, which includes two sets of core attributes for the "Requirements" concept and "Product Backlog" concept (Table 1).

- Note that a concept in linguistics is assumed as a noun or noun-phrase, which serves as the subject of a to-be statement [8]. By using a Linguistic Typological Analysis (LTA) (assuming that a simple sentence is made of "subject", "predicate" and "object" parts), an initial similarity level is determined on a scale ranging from 0 to 3.

- "0" level indicates no-typological similarity where none of the parts of two attributes is similar. "1" indicates that one similar part exists. "2" means that two of linguistic parts are similar. Finally, 3 points out a full linguistic similarity where both of the sentences have similar

"subject", "predicate" and "object" parts. Note that the level 2 or 3 is regarded as satisfactory for EF mapping procedures in this study.

- By using the definition (4), $A_{EF} \cap A_{SEP} \neq \emptyset$, we find that two concepts share some common attributes ($a_3$-$b_3$; $a_4$-$b_4$; $a_6$-$b_6$). LTA also shows that these three attributes have a linguistic similarity level at 2 (Table 1)

Table 1. Requirements and Product Backlog Attribute Sets

| Set of attributes for "Requirements" concept of EF | Set of attributes for "Product Backlog" concept of Scrum | Linguistic similarity level (0 to 3) |
|---|---|---|
| $A_{EF} = \{a_1, a_2, \ldots, a_n\}$ | $A_{SEP} = \{b_1, b_2, \ldots, b_n\}$ | |
| $a_1$ = are the definition of what needs to be achieved | $b_1$ = is a prioritized list of desired product functionality | 1 |
| $a_2$ = must address opportunity and satisfy stakeholders | $b_2$ = is required to meet the product owner's vision | 1 |
| $a_3$ = mechanisms for managing /accepting requirements need to be established | $b_3$ = product owner is responsible for determining and managing requirements | 2 |
| $a_4$ = progress through six states: conceived, bounded, coherent, acceptable, addressed, fulfilled | $b_4$ = the definition of ready and the definition of done are two major states of product backlog items (PBIs) | 2 |
| $a_5$ = must be bounded as a whole and stay within the bounds of original concept | $b_5$ = provides shared understanding of (a) what to build and (b) the order of what to build. | 1 |
| $a_6$ = continue to evolve as more is learned. | $b_6$ = Grooming is important and it refers to creating, refining, estimating and prioritizing PBIs continually. | 2 |

- By using the definition (5):

$$\frac{\#(\{a1,a2,a3,a4,a5,a6,\} \cap \{b1,b2,b3,b4,b5,b6\})}{\#(\{a1,a2,a3,a4,a5,a6\} \cup \{b1,b2,b3,b4,b5,b6\})} * 100$$

The conceptual similarity level is consequently found as

$$= \frac{3}{9} * 100 = 33.3$$

This finding indicates a result, which may be regarded as different from the specifications or Essence-based Scrum practice definitions mentioned in the Essence Literature. Such that the "Requirements" and "Product Backlog" concepts are not conceptually equal as it is claimed or specified.

At first glance, the most of experts on both Scrum and EF may not object to association of "Requirements" and "Product Backlog". However, the result is substantially different in our sample case. It is thought that the primary reason would be the human experience and informal judgement, which is usually adopted in mapping procedures in the literature. For example, in [4] and [5], an ontology of terms, commitments and metamodeling techniques guide the mapping processes. However, their classifications of SE practice terms into a list of corresponding EF concepts, such as, work products, activities, roles, which again employ subjective expert judgements. In another study proposing an algorithm [4], the assignment of SE practice activities to EF activity spaces, specifying their alpha states and checklists are also dependent on personal experience and subjective expert evaluations.

## 5. CONCLUSIONS

Concepts are important for carrying certain meanings in thinking, reasoning and system modelling [8]. By using CA, SE practices and EF can be modelled as dynamic and abstract mathematical structures that encapsulate objects as well as their attributes and relations. This study shows that CA can provide the formal and generic knowledge manipulation means required for complex software and knowledge structures.

As a relatively new framework proposed for the core problems of SE methods, one important issue for EF has been the mapping a SE practice to the EF's conceptual domain. Thus, the main argument of this paper is that formal methods can provide more accurate transformations as well as they can enable application of more systematic mapping procedures. In this study, therefore, a formal method using an algorithm and CA definitions is proposed as a complete solution. This is applied in a Scrum case and the results differ from the ones exist in the current EF literature. However, the research limitations confine us within presenting mainly foundations, a sample case and its initial observations. More empirical evidences from software industry are needed to support the mapping method. Therefore, the paper concludes with an invitation to future studies aiming to address the research limitations.

International Journal of Software Engineering & Applications (IJSEA), Vol.9, No.6, November 2018

## AUTHORS

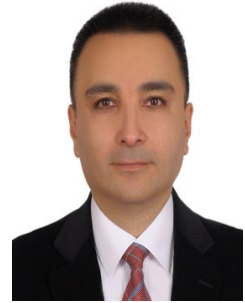

**Assoc. Prof. Dr. Murat Pasa Uysal**, is at the Department of Management Information Systems in Baskent University. He holds a B.S degree in electrical & electronic engineering, a M.S degree in computer engineering, and a Ph.D. degree in educational technology. He completed his post-doctoral studies at Rochester Institute of Technology in New York, which was on software re-engineering and Information Technologies (IT) Governance. He served as an advisor and engineer for different types of IT projects in Turkish Army (TA) for many years, and conducted studies addressing the problems of TA in the research areas of IT. He has been teaching IT, information systems and software engineering related courses. His research interest is also in the areas of software engineering, information systems (IS), instructional methods and tools for computer programming and IS.